\documentclass{aa}
\usepackage[dvips]{graphics}
\usepackage{times}

\begin{document}

\thesaurus{05(10.15.2 NGC 3532; 08.03.5; 13.25.5)}

\title{A $ROSAT$ HRI study of the open cluster NGC~3532}

\author{E. Franciosini\inst{1} \and S. Randich\inst{2} \and R. 
Pallavicini\inst{1}}

\institute{
Osservatorio Astronomico di Palermo ``G.S. Vaiana'', Piazza del Parlamento
1, I-90134 Palermo, Italy (francio@oapa.astropa.unipa.it,
pallavic@oapa.astropa.unipa.it)
\and
Osservatorio Astrofisico di Arcetri, Largo E. Fermi 5, I-50125 Firenze, Italy
(randich@arcetri.astro.it)
}

\offprints{E. Franciosini}

\date{Received  / Accepted  }

\maketitle
\begin{abstract}
NGC~3532 is a very rich southern open cluster of age $\sim 200-350$ Myr; it
is therefore a good candidate to investigate the X-ray
activity--age--rotation relationship at ages intermediate between the
Pleiades and the Hyades, where, to our knowledge, X-rays studies exist for
only one cluster (NGC~6475).
We have performed an X-ray study of NGC~3532 using HRI observations
retrieved from the $ROSAT$ archive. The observations have a limiting
sensitivity $L_{\mathrm x} \sim 4 \times 10^{28}$ erg~sec$^{-1}$ in the
center of the field. We detected $\sim 50$ X-ray sources above a $4 \sigma$
threshold, half of which have a known optical counterpart within 10 arcsec;
15 of the X-ray sources have at least one cluster member as optical
counterpart.

A comparison of NGC~3532 with the nearly coeval cluster NGC~6475 indicates
that the former cluster is considerably X-ray underluminous with respect to
NGC~6475. However, because of the existence of possible selection effects,
additional X-ray and optical observations are needed before definitively
concluding that the X-ray properties of NGC~3532 and NGC~6475 are
significantly different.

\keywords{open clusters and associations: individual: NGC~3532 -- stars:
coronae -- X-ray: stars}
\end{abstract}

\section{Introduction}

The $ROSAT$ PSPC and HRI detectors have provided X-ray images for a large
number of open clusters sampling the age range from $\sim$~20 to 600~Myr
(e.g., Randich \cite{randich00} and references therein; Jeffries
\cite{jeffries99} and references therein; see also Belloni \cite{belloni97},
for a review on older open clusters). The data have allowed investigating in
great detail the dependence of X-ray activity on  mass, age, rotation, and,
in particular, to check the validity of the rotation--activity--age
paradigm. The overall picture emerging from $ROSAT$ generally confirms that
there is a tight dependence of X-ray activity on rotation (or on the so
called Rossby number, the ratio of the rotation period over the convective
turnover time -- e.g., Noyes et al. \cite{noyes84}) and, through rotation,
on age: the level of X-ray activity increases with increasing rotation and,
since stars spin down as they age, the average or median X-ray luminosity
decays with increasing age. However, the X-ray luminosity (or X-ray over
bolometric luminosity) does not depend simply on some power of the
rotational rate, and the activity--age dependence cannot be described by a
Skumanich--type power law. In addition, a few puzzling results have arisen
from $ROSAT$ data.
For example, the finding that the bulk of the population of Praesepe
solar-type stars have a significantly lower X-ray luminosity than the coeval
Hyades and Coma Berenices clusters (Randich \& Schmitt \cite{randich95};
Randich et al. \cite{randich96}) has casted doubts on the common thinking
that a unique activity--age relationship holds, and, consequently, that the
X-ray properties of a cluster of a given age are representative of all
clusters of the same age. A study by Barrado y Navascu\'es et al.
(\cite{barrado98}) seems to exclude  that this result is due to a strong
contamination of the Praesepe sample by cluster non-members; at the same
time, $ROSAT$ observations of NGC~6633 suggest that this cluster, which is
coeval to the Hyades and Praesepe, is more Praesepe--like than Hyades--like
(Franciosini et al. \cite{francio00}; Totten et al. \cite{totten00}). We
also mention that the comparison of the Pleiades (120~Myr) with NGC~6475
(200~Myr) and with other clusters with ages of the order of 100--200 Myr
also suggests that a tight/unique age--activity relationship may not hold
(e.g. Randich \cite{randich00}). The issue of the uniqueness of the
activity--age relationship is therefore not at all settled. In addition to
optical studies that should ascertain cluster membership and provide
complete (or close to completeness) lists of members and better defined
cluster ages, additional, and possibly deeper, X-ray surveys of samples of
coeval clusters are clearly required to further address this problem.

We present here a $ROSAT$ study of the NGC~3532 cluster: NGC~3532 is a very
rich southern open cluster with an estimated age $\sim 200-350$ Myr
(Fernandez \& Salgado \cite{fs80}; Johansson \cite{johan81}; Eggen
\cite{eggen81}; Koester \& Reimers \cite{koester93}; Meynet et al.
\cite{meynet93}); it is therefore a good candidate to investigate the X-ray
activity--age--rotation relationship at ages intermediate between the
Pleiades and the Hyades, where, to our knowledge, X-rays studies exist for
only one cluster (NGC~6475). 
The most likely value for the reddening of NGC~3532 is $E(B-V) = 0.04$
(Fernandez \& Salgado \cite{fs80}; Eggen \cite{eggen81}; Schneider
\cite{schneid87}; Meynet et al. \cite{meynet93}); the metallicity of the
cluster has been estimated to be close to solar ([Fe/H] $\sim 0.02$;
Clari\'a \& Lapasset \cite{claria88}). The cluster is located at very low
galactic latitude ($b=+1.43$ ~deg). Distance determinations range from
405$^{+76}_{-55}$~pc (from Hipparcos; Robichon et al. \cite{robich99}) to
500 pc (Eggen \cite{eggen81}); in this paper the most recent value of 405 pc
by Robichon et al. (\cite{robich99}) has been adopted.

\begin{figure}
\resizebox{\hsize}{!}{\includegraphics{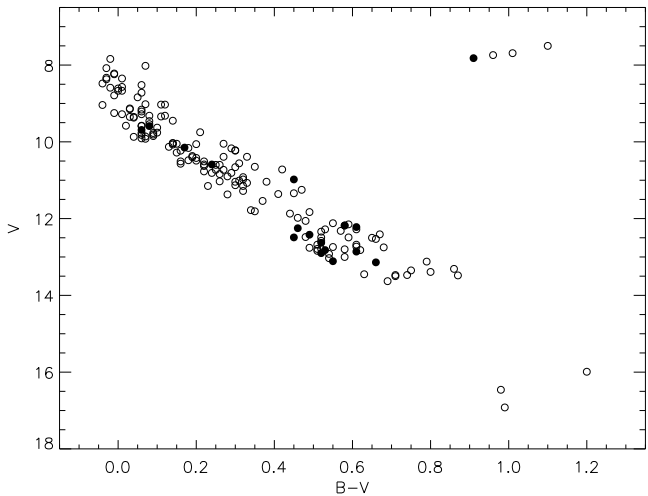}}
\caption{
$V$ vs. $(B-V)$ diagram for the probable and possible members of NGC~3532
included in the HRI field of view. Filled circles indicate stars detected in
X-rays
}
\label{cmd}
\end{figure}

\section{Optical catalog}

The first detailed study of NGC~3532 was carried out by Koelbloed
(\cite{koelb59}), who obtained photoelectric or photographic photometry and
proper motions for 255 stars down to a limiting magnitude $V \sim 11.7$. A
new proper motion survey of these stars was later performed by King
(\cite{king78}). The most extensive study of this cluster is the photometric
study by Fernandez \& Salgado (\cite{fs80}), who obtained photoelectric and
photographic photometry for 700 stars (including nearly all Koelbloed's
stars) down to a limiting magnitude $V = 13.5$. Photoelectric photometry for
another 24 stars down to $V = 18.3$ was obtained by Butler
(\cite{butler77}). We mention in passing that only 15 G--type and 7 K--type
dwarf cluster members are present in the total sample of 724 stars.
Additional photometric studies of these stars have been performed by
Johansson (\cite{johan81}; UBV, 16 stars), Eggen (\cite{eggen81};
Str\"omgren, 33 stars), Wizinowich \& Garrison (\cite{wiz82}; UBVRI, 68
stars), Schneider (\cite{schneid87}; Str\"omgren, 164 stars) and Clari\'a \&
Lapasset (\cite{claria88}; UBV and DDO, 12 stars). Radial velocities are
available for about a hundred stars from the studies by Harris
(\cite{harris76}) and by Gieseking (\cite{gies80}, \cite{gies81}). Gieseking
(\cite{gies81}) derived a mean cluster radial velocity $v_{\rm r} = 4.6 \pm
2$ km/s.

Our input catalog is based on the lists of stars by Fernandez \& Salgado
(\cite{fs80}) and Butler (\cite{butler77}). From these lists, we selected as
{\it probable} members those stars with radial velocity, when available, 
within 4 km/s (i.e. $2\sigma$) of the cluster mean $v_{\rm r}$, or with
membership probability from proper motions greater than 80\%, or which were
suggested as members in photometric studies. We rejected stars that would be
considered members according to either radial velocity or proper motion, but
with photometry inconsistent with cluster membership. For stars with no
individual membership information, but with UBV photometry available, we
accepted as {\it possible} members those falling in a band between $0.2^m$
below and $0.7^m$ above the cluster main sequence. 

The resulting catalog contains 248 probable and possible members; 174 of
them, including 4 giants, are located within 17 arcmin of the $ROSAT$
nominal pointing position. In Fig.~\ref{cmd} we show the $V$ vs. $(B-V)$
C--M diagram for the probable and possible members in our field of view. It
is evident from the figure that the majority of the known members are
early-type stars. Except for three very late possible cluster members, the
cluster main sequence is truncated at $V= 13.5$, corresponding to G--type
stars; only 13 G--type and 5 K--type members (excluding giants) are present in
our catalog, compared to 104 B--A and 48 F stars. We also mention that most
of the stars with spectral type later than F5 were selected as members only
on the basis of photometry.

\begin{table*}
\caption{Detected X-ray sources identified with cluster members within
10$\arcsec$. Star numbering for cluster members is from Fernandez \& Salgado
(\cite{fs80})} 
\begin{tabular}{rrrrrrlccrrl}
\hline
No.& $\alpha_{\rm x}$ (2000)& $\delta_{\rm x}$ (2000)& ML& count rate & 
$L_{\mathrm x}$\ \ \ \ \ \ \ \quad & Optical& $\Delta r$\hfil & Memb.& V & 
B-V & Notes \\
 & & & & ($10^{-5}$ s$^{-1}$)& ($10^{29}$ erg/s)& ident. & ($\;\arcsec\;$) & 
 pm $v_r$ ph& & & \\
\hline
 4& 11 05 53.07& -58 35 34.2&  14.6&  45 $\pm$   10&  2.3 $\pm$ 0.5& FS243& 
4.9& y - y& 10.98&  0.45&\\
 5& 11 05 46.99& -58 36 47.3&  22.0&  40 $\pm$\ \ 8&  2.0 $\pm$ 0.4& FS242& 
2.1& y - y& 12.18&  0.58&\\
 8& 11 06 33.24& -58 37 45.7&  32.9&  67 $\pm$   11&  3.4 $\pm$ 0.6& FS229& 
3.9& - - y& 13.14&  0.66&\\
11& 11 04 37.02& -58 39 05.5&  17.8&  57 $\pm$   12&  2.9 $\pm$ 0.6& FS146& 
4.1& - - y& 12.63&  0.52&\\
12& 11 05 33.78& -58 39 08.9&  12.7&  31 $\pm$\ \ 8&  1.6 $\pm$ 0.4& FS128& 
2.6& y - y& 12.42&  0.49&\\
18& 11 05 35.15& -58 40 37.5&  12.7&  31 $\pm$\ \ 8&  1.5 $\pm$ 0.4& FS129& 
0.8& y - y& 10.59&  0.24&\\
21& 11 04 52.84& -58 40 53.9&  37.3&  78 $\pm$   12&  3.9 $\pm$ 0.6& FS149& 
2.4& - - y& 12.86&  0.61&\\
22& 11 06 23.06& -58 40 59.3&  13.7&  27 $\pm$\ \ 7&  1.4 $\pm$ 0.4& FS102& 
2.4& - - y& 12.90&  0.52&\\
  &            &            &      &               &               & FS103& 
7.8& - - y& 12.49&  0.45&\\
25& 11 04 33.38& -58 41 37.1&  25.4&  66 $\pm$   12&  3.4 $\pm$ 0.6& FS152& 
5.2& y ? y&  7.82&  0.91& g, SB?\\
27& 11 05 49.99& -58 42 18.1&  10.2&  28 $\pm$\ \ 8&  1.4 $\pm$ 0.4&  FS21& 
5.9& y - y&  9.59&  0.08&\\
37& 11 05 18.25& -58 46 16.4&  27.7&  39 $\pm$\ \ 8&  2.0 $\pm$ 0.4&   FS8&
2.4& y - y&  9.69&  0.06&\\
40& 11 04 35.32& -58 48 26.6&  60.7& 127 $\pm$   16&  6.5 $\pm$ 0.8& FS169&
2.6& - - y& 12.25&  0.46&\\
  &            &            &      &               &               & FS170&
8.7& y - y& 10.15&  0.17&\\
42& 11 05 28.75& -58 49 29.3&  12.6&  37 $\pm$\ \ 9&  1.9 $\pm$ 0.5&  FS57&
1.7& - - y& 13.11&  0.55&\\
48& 11 04 14.02& -58 45 52.6&  11.2&  69 $\pm$   17&  3.5 $\pm$ 0.9& FS277&
0.0& - - y& 12.82&  0.53&\\
49& 11 05 52.18& -58 55 35.5&  10.5&  63 $\pm$   16&  3.2 $\pm$ 0.8& FS314&
0.0& y - y& 12.22&  0.61&\\
\hline
\noalign{\smallskip}
\multicolumn{12}{l}{{\bf Note}: for proper motion membership, `y' means $P
\ge 80 \%$, `?' means $65\% \le P <80\%$}\\
\multicolumn{12}{l}{A `-' in the membership columns indicates that no
information is available}\\
\end{tabular}
\label{xmembers}
\end{table*}

\section{Observations and data analysis}

The X-ray data used in this study have been retrieved from the $ROSAT$ public
archive (obs. IDs 202075h, 202075h-1, 202075h-2). NGC~3532 was observed with
the HRI during three separate pointings on January 21, 1996, July 28, 1996,
and June 19, 1997. The net exposure times were respectively 30.5 ksec, 37
ksec, and 34 ksec. The nominal pointing position for all observations was RA
$= 11^{\rm h} 5^{\rm m} 43.2^{\rm s}$, DEC $= - 58\degr 43\arcmin 12\arcsec$
(J2000).

The analysis was performed using EXSAS routines within MIDAS. We first
checked the alignment of the three single images by comparing the positions
of common sources; since the shifts between the images are very small (less
than 1 image pixel), we did not apply any correction to the data.
The three Photon Events Tables (PET) were then merged into a single PET,
from which an image with a total exposure time of 101.5 ksec was generated.
We then followed the standard steps for data reduction. A background map was
created from the global image by removing outstanding sources previously
detected with the {\sc local/detection} algorithm and then smoothing with a
spline filter. Source detection was performed using the Maximum Likelihood
(ML) algorithm. The ML algorithm was first run on a provisional list of
sources obtained from the Local and Map Detection, resulting in the
detection of 47 sources with ML $>10$ (corresponding to a significance of
4$\sigma$), lying within 17 arcmin from the image center; two additional
sources (nos. 48 and 49) were detected above the same threshold by running
the ML on the input optical catalog. Of these sources, 15 have at least one
cluster member counterpart within 10 arcsec, 13 have an optical counterpart
which is probably a cluster non-member, and 21 do not have any known optical
counterpart (additional positions of non-member stars from the survey of
Andersen \& Reiz \cite{anders83} and from the Guide Star Catalog have also
been considered). The X-ray and optical properties of the sources with an
optical counterpart are listed in Tables~\ref{xmembers} (cluster members)
and \ref{xident} (non-members); the list of unidentified sources is given in
Table~\ref{unident}. For the cluster members without associated X-ray
sources we estimated 3$\sigma$ upper limits from the background count rates
at the optical position. 

We note that sources no. 27 in Table~\ref{xmembers} and nos. 24 and 31 in
Table~\ref{xident} are barely visible above the background on the X--ray
image (as indicated also by their low ML) and therefore could be not real.
However, since two of them are identified with cluster non-members (nos. 24
and 31) and the other with an A-type cluster member (no. 27), including or
excluding them from our source list would not change our main
results/conclusions.

We estimated the number of spurious identifications due to chance
coincidences, following Randich et al. (\cite{randich95a}). Such a number
($N_{\rm s}$), is given by:
\begin{equation}
N_{\rm s}=D_{\rm c} \times N_{\rm X} \times A_{\rm id.}
\end{equation}
where $D_{\rm c}$ is the density of cluster candidates within the surveyed
area (i.e., the number of clusters candidates divided by the HRI field of
view), $N_{\rm X}$ is the number of X--ray sources, and $A_{\rm id.}$ is the
area of our identification circle. Considering $D_{\rm c}=174/(289\times
\pi)$~arcmin$^{-2}$, $N_{\rm X}=49$, and $A_{\rm id.}= 0.028\times
\pi$~arcmin$^2$, we obtain $N_{\rm s}=0.83$, i.e., less than one spurious
identification.

\begin{table*}
\caption{Detected X-ray sources with an optical counterpart within
10$\arcsec$ which is probably a cluster non-member}
\begin{tabular}{rrrrrlccrrl}
\hline
No.& $\alpha_{\rm x}$ (2000)& $\delta_{\rm x}$ (2000)& ML& count rate & 
Optical& $\Delta r$\hfil & Memb.& V & B-V & Notes\\
 & & & & ($10^{-5}$ s$^{-1}$)& ident. & ($\;\arcsec\;$) & 
 pm $v_r$ ph& & & \\
\hline
 1& 11 06 16.23& -58 33 32.9& 259.8& 307 $\pm$   21& FS388& 2.9& n - n&
11.52&  0.76& \\
 7& 11 05 03.49& -58 37 27.6&  64.6& 101 $\pm$   13& GSC8627-2833& 3.6& - - -& 
12.46& & \\
13& 11 05 22.34& -58 39 10.9&  15.1&  28 $\pm$\ \ 7& FS132& 1.1& n - n& 
11.30&  0.53& \\
15& 11 04 17.75& -58 39 42.5&  34.1& 112 $\pm$   17& FS268& 0.9& - - n&
13.22&  1.02& \\
16& 11 05 53.74& -58 39 45.8&  25.5&  42 $\pm$\ \ 8& FS115& 1.6& n - y&
8.51& -0.11& \\
19& 11 05 45.68& -58 40 38.1& 884.2& 504 $\pm$   24& FS122& 1.8& ? - n&
8.20&  0.93& {\it a} \\
24& 11 06 45.78& -58 41 22.1&  11.7&  35 $\pm$\ \ 9& FS219& 2.1& n - n&
11.23&  0.53& \\
26& 11 05 06.37& -58 42 09.4&  20.6&  44 $\pm$\ \ 9&  FS35& 0.5& - - n&
13.36&  0.42& \\
31& 11 06 08.79& -58 43 27.3&  10.9&  20 $\pm$\ \ 6&  FS90& 1.0& n - n&
11.78&  0.63& \\
33& 11 06 52.76& -58 44 58.3&  89.3& 130 $\pm$   14& FS354& 1.3& n - n&
12.42&  0.34& \\
36& 11 06 08.98& -58 45 52.4&  36.2&  47 $\pm$\ \ 8&  FS84& 0.9& n - y&
12.03&  0.53& \\
41& 11 05 54.92& -58 49 02.0&  97.0& 111 $\pm$   13&  FS67& 1.9& n - y&
9.81&  0.16& \\
43& 11 06 01.14& -58 50 26.6&  27.5&  52 $\pm$   10& FS196& 1.7& y - n&
11.96&  0.74& \\
\hline
\noalign{\smallskip}
\multicolumn{11}{l}{
{\it a)} extended source. No known optical extended sources are present at
the X-ray position
}\\
\end{tabular}
\label{xident}
\end{table*}

X--ray luminosities were derived as follows. We assumed a conversion factor
(CF) of $2.6 \times 10^{-11}$ erg~cm$^{-2}$~sec$^{-1}$ per HRI
count~sec$^{-1}$,  estimated using PIMMS (version 2.7) assuming a
Raymond-Smith plasma with $T = 10^6$ K and a column density $\log N_H =
20.3$ cm$^{-2}$; higher temperatures do not significantly affect the value
of the conversion factor, and the same is true if a two-temperature model is
assumed. X-ray luminosities for both detections and upper limits were then
computed assuming a cluster distance of 405 pc. The resulting sensitivity in
the center of the field is $L_{\mathrm x} \sim 3.6 \times 10^{28}$
erg~sec$^{-1}$, a factor $\sim$ 2 higher than the limiting sensitivity of
the X-ray studies of the coeval cluster NGC~6475 (Prosser et al.
\cite{prosser95}; James \& Jeffries \cite{james97}). Had we assumed a 10\%
larger distance to the cluster ($d = 450$ pc), the X-ray luminosities and
upper limits would have been a factor of 20\% larger, not introducing any
significant change in our results.
Note that, due to the relative short exposure times of the three individual
images,  we are not able to put stringent constraints on source variability.
We just mention that for the few X-ray sources that were detected in the
single images we obtained count rates very similar to the ones that we
inferred from the global image.

\begin{table}
\caption{Unidentified X-ray sources}
\begin{tabular}{rrrrrlccrrl}
\hline
No.& $\alpha_{\rm x}$ (2000)& $\delta_{\rm x}$ (2000)& ML& count rate \\
 & & & & ($10^{-5}$ s$^{-1}$) \\
\hline
 2& 11 06 03.56& -58 35 08.8&  27.7&  60 $\pm$  11 \\
 3& 11 05 42.97& -58 35 23.6&  39.0&  71 $\pm$  11 \\
 6& 11 06 29.65& -58 37 07.4& 195.9& 203 $\pm$  17 \\
 9& 11 05 09.88& -58 38 29.2&  19.5&  38 $\pm$ \ 8 \\
10& 11 05 02.23& -58 38 58.9&  12.1&  34 $\pm$ \ 9 \\
14& 11 06 05.99& -58 39 35.7&  14.8&  30 $\pm$ \ 7 \\
17& 11 05 25.66& -58 40 09.0&  13.0&  26 $\pm$ \ 7 \\
20& 11 06 19.93& -58 40 46.7&  34.7&  46 $\pm$ \ 8 \\
23& 11 06 18.71& -58 41 03.0&  14.8&  33 $\pm$ \ 8 \\
28& 11 05 12.42& -58 42 18.5&  28.6&  44 $\pm$ \ 8 \\
29& 11 06 54.20& -58 42 23.0&  13.7&  53 $\pm$  12 \\
30& 11 05 52.48& -58 42 59.0&  32.8&  51 $\pm$ \ 9 \\
32& 11 05 06.18& -58 44 42.9&  11.8&  36 $\pm$ \ 9 \\
34& 11 04 50.82& -58 45 46.1&  17.8&  46 $\pm$  10 \\
35& 11 07 03.09& -58 45 50.1&  13.2&  55 $\pm$  13 \\
38& 11 05 45.86& -58 47 29.0&  10.5&  21 $\pm$ \ 6 \\
39& 11 05 20.57& -58 47 58.3&  31.9&  64 $\pm$  11 \\
44& 11 05 10.28& -58 50 33.0&  16.1&  43 $\pm$  10 \\
45& 11 05 43.34& -58 50 55.6&  32.7&  68 $\pm$  11 \\
46& 11 05 52.01& -58 51 11.7&  21.6&  53 $\pm$  11 \\
47& 11 05 53.15& -58 54 24.6&  14.9&  63 $\pm$  14 \\
\hline
\end{tabular}
\label{unident}
\end{table}

\section{Results}

As mentioned in the previous section, 15 sources have been identified with
cluster members. For two sources (nos. 22 and 40) two cluster members are
found within the identification radius. Our analysis resulted in the
detection of 11 F--type cluster stars out of 48 (detection rate 23\%), one
G--type dwarf out of 13 (detection rate 8\%), and one of the four giants.
None of the five K dwarfs in our field has been detected. Four A--type stars
were also detected. The detected stars are indicated as filled symbols in
Fig.~\ref{cmd}. 
The issue of X-ray emission from early-type (i.e., earlier than F0) stars
which, due to the lack of a convective zone, cannot generate magnetic fields
(and thus magnetic activity) via the dynamo process, has been discussed at
length in several papers (e.g., Micela et al. \cite{micela96} and references
therein); the most likely possibility is that their X-ray emission is due to
unseen binary companions. Therefore, we focus the following discussion on
solar-type (namely, F and G-type) stars only.

As to the X-ray sources identified with non-members, they do not warrant
much further discussion. Most of them, as indicated by their position on the
C--M diagram are most likely G/early--K type foreground stars. Given that the
cluster is basically located on the galactic plane, it is not surprising 
to find such a large contamination from cluster non-members among X-ray
sources.

\subsection{Comparison with other clusters}

\begin{figure}
\resizebox{\hsize}{!}{\includegraphics{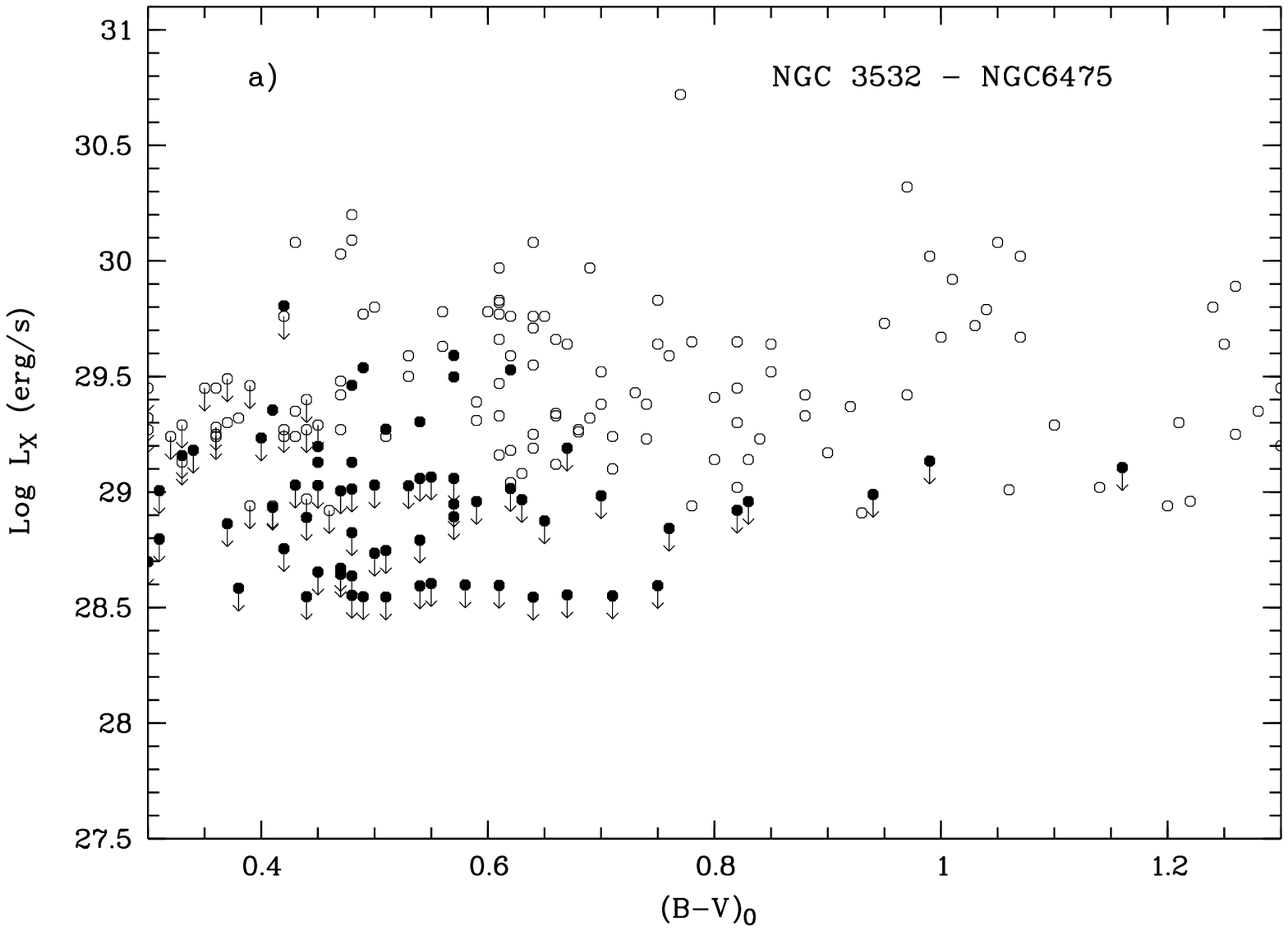}}
\resizebox{\hsize}{!}{\includegraphics{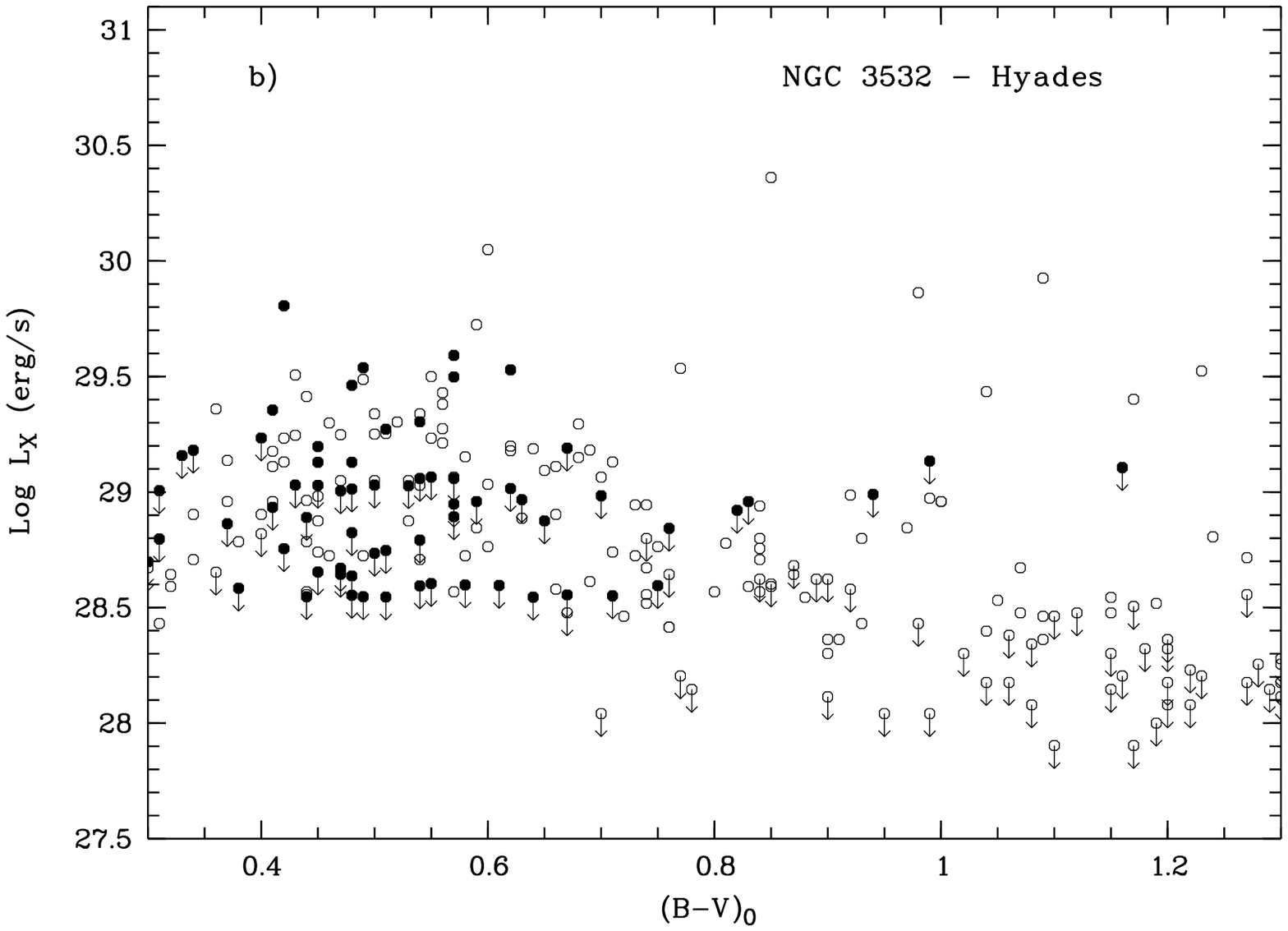}}
\caption{
Comparison of the relation $\log L_{\mathrm x}$ vs. $(B-V)_0$ of NGC~3532
with NGC~6475 (panel a) and the Hyades (panel b). Filled symbols denote
NGC~3532 members, while open symbols indicate NGC~6475 and Hyades stars
}
\label{lxbv}
\end{figure}

In Figs.~\ref{lxbv}a--\ref{lxbv}b we compare the $\log L_{\mathrm x}$ vs.
$(B-V)_0$ distribution of NGC~3532 with those of the supposedly coeval
NGC~6475 cluster and the older Hyades. The comparison with NGC~6475
(Fig.~\ref{lxbv}a) suggests that the bulk of NGC~3532 F and G-type stars may
be less X-ray luminous than NGC~6475. The few detections have X-ray
luminosities comparable to the luminosities of similar stars in NGC~6475,
but the majority of NGC~3532 solar-type stars were not detected; most
important, the upper limits we derived for a very large fraction of the
late--F and G--type stars in NGC~3532 are as low as or even below the
luminosities of the least X-ray luminous stars of NGC~6475.

\begin{figure}[t]
\resizebox{\hsize}{!}{\includegraphics{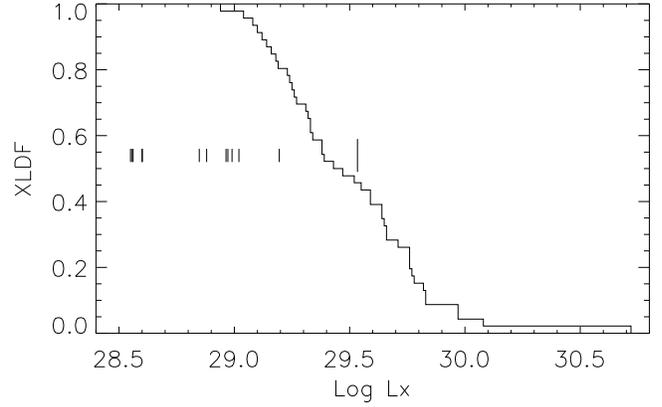}}
\caption{
Comparison of the X-ray luminosity distribution function (XLDF) for G dwarfs
($0.59 \leq (B-V)_0 < 0.81$) in NGC~6475 (solid curve) with the upper limits
(short vertical bars) and the one detection (long vertical bar) derived for
NGC~3532. The XLDF for NGC~6475 has been constructed using the data from
Prosser et al. (\cite{prosser95}) and James \& Jeffries (\cite{james97})
}
\label{xldf}
\end{figure}

Given the low number of detections, a direct comparison of the X-ray
luminosity distribution function (XLDF) of NGC~3532 with the XLDF of the
coeval cluster NGC~6475 would not be of much help. In Fig.~\ref{xldf} we
show instead the X-ray luminosity distribution function (XLDF) for G--type
stars with $0.59 \leq (B-V)_0 < 0.81$ in NGC~6475 with vertical bars
indicating the upper limits and the one detection in this spectral range for
NGC~3532. The figure seems to confirm that the population of solar-type
stars in NGC~3532 is less X-ray active than NGC~6475. Such a conclusion is
supported by a statistical comparison of the X-ray properties of G dwarfs in
the two clusters, carried out using various two-sample tests as implemented
in the Astronomy SURVival Analysis ({\sc asurv}) Ver. 1.2 software package 
(see Feigelson \& Nelson \cite{feig85}; Isobe et al. \cite{isobe86}); 
the tests indicate that the hypothesis that NGC~3532 and NGC~6475 solar-type
stars are drawn from the same parent population can be rejected with a
confidence level higher than 99.9 \%. In addition, considering the XLDF of
NGC~6475 and using the method described by Randich et al. (\cite{randich98})
for IC~4756, we estimate that the probability of getting the observed ULs
distribution of NGC~3532 if the XLDF of NGC~3532 were the same as the one of
NGC~6475 is virtually 0.

Several possibilities can explain our results: 
{\bf a)} first, and most obviously, the reddening to the cluster could be
significantly wrong; a higher reddening would mean a higher column density
of absorbing material and would eventually imply that our upper limits (as
well as the X--ray luminosities of the detected stars) are underestimated.
However, all the sources in the literature, using different methods, agree
in deriving a reddening to the cluster $E(B-V)\leq 0.1$, with the most
quoted value being in fact $E(B-V)=0.04$. If we assume a reddening as high
as $E(B-V)=0.1$ (Johannson et al. \cite{johan81}), we get a factor 1.5
higher CF for $T=10^6$~K (CF~$= 4.0\times 10^{-11}$ instead of $2.6 \times
10^{-11}$ erg~cm$^{-2}$~sec$^{-1}$ per HRI count~sec$^{-1}$) and the same CF
for higher temperatures; similar results are found using two-temperature
models. Therefore, it seems rather unlikely that the use of an incorrect
value for the reddening is the major cause of the discrepancy between
NGC~6475 and NGC~3532;
{\bf b)} second, NGC~6475 is an X-ray selected sample, i.e. most of its
solar--type and lower mass members were not known until X--ray surveys of
the cluster were carried out and they were detected in X-rays. Therefore, we
cannot exclude that a low activity (with X-ray luminosities below 10$^{29}$
erg~sec$^{-1}$ -- see Fig.~\ref{lxbv}a) population exists that was not
detected in the two $ROSAT$ surveys of this cluster. The comparison of the
XLDF of NGC~6475 with that of the Pleiades or other young clusters indeed
suggests that this is a very likely possibility. Such a population would
contribute to the low luminosity tail of NGC~6475 distribution function;
nevertheless, Fig.~\ref{lxbv}a indicates that, as a matter of fact, NGC~3532
also lacks the high luminosity population that is present in NGC~6475. We
conclude that, although the presence of an X-ray faint population in
NGC~6475 would partly reduce the inconsistency between the two clusters, it
could not completely cancel it, unless one assumes that the low X-ray
luminosity population of NGC~6475 is 5--10 times more numerous than the high
luminosity one; 
{\bf c)} the NGC~3532 sample is incomplete and the membership for most of
the late-type cluster members is based on photometry only. Therefore, on the
one hand, our optical sample could be highly contaminated by non-members
and, on the other hand, several other optically unknown members could exist.
If all or most of the 21 X-ray sources without a known optical counterpart
turn out to be solar-type (or later) cluster members and, at the same time,
part of the optically selected members turn out to be non-members, the
discrepancy between NGC~6475 and NGC~3532 would possibly be solved. The 21
unidentified X-ray sources if located at the cluster distance would have
X-ray luminosities in the range $1.1 \times 10^{29} - 1.0 \times 10^{30}$
erg~sec$^{-1}$; if all these sources were G--type cluster members, the XLDF
for NGC~3532 would have indeed a median $\log L_{\mathrm x} = 29.3$,
slightly lower than the median for NGC~6475 (29.4). Therefore we cannot
exclude that the results presented here are due, at least in part, to the
incompleteness of the presently known optical cluster sample; nevertheless,
if this were true, it would be difficult in any case to explain why virtually
all the currently known solar-type cluster members are X-ray faint;
{\bf d)} if neither point {\bf b)} or {\bf c)} (or both together) were
proven to explain entirely why NGC~3532 is less X-ray luminous than
NGC~6475, then the conclusion could be drawn that there is a {\it real}
difference between the X-ray properties of the two clusters. In this case,
two hypothesis could be made: {\it i)} NGC~3532 is actually older than
NGC~6475; {\it ii)} NGC~6475 and NGC~3532 are about coeval, and our result
represents an additional piece of evidence that the age--activity
relationship is not unique. Fig.~\ref{lxbv}b indeed indicates that the X-ray
properties of NGC~3532 may be more similar to those of the Hyades than to
NGC~6475. Using again the two sample tests, we find that the hypothesis that
NGC~3532 and Hyades solar-type stars are drawn from the same parent
population can be excluded with a confidence level ranging between 95 and 98
\%, depending on the adopted test. We mention that the age of NGC~3532 has
been generally estimated using C--M diagram fitting or, in two cases, from
the magnitude of the turn-off. As mentioned in the introduction different
methods result in an age between 200 Myr (Fernandez \& Salgado \cite{fs80};
Johansson \cite{johan81}) and 350 Myr (Eggen \cite{eggen81}); the most
recent determinations give $\sim 300$ Myr (Koester \& Reimers
\cite{koester93}; Meynet et al. \cite{meynet93}). Note that Meynet et al.
(\cite{meynet93}) using the same method/isochrones derived an age of $\sim
220$ Myr for NGC~6475; it seems, therefore, that NGC~3532 might be slightly
older than NGC~6475, but not as old as the X-ray data would suggest.

\section{Conclusions}

We have analyzed $ROSAT$ archive data of the open cluster NGC~3532. The
comparison of the X-ray properties of solar-type stars in the cluster with
those of the supposedly coeval NGC~6475 cluster indicates that NGC~3532 is
considerably X-ray underluminous with respect to NGC~6475. If this result
is not due to selection effects and biases in the two cluster samples, it
would provide an additional piece of evidence that the X-ray activity--age
relationship is not unique and that other parameters, in addition to
rotation, determine the level of coronal emission. However, before such a
conclusion can be accepted, additional X-ray and optical observations should
be performed. Namely, {\bf I.} an additional X-ray survey of NGC~6475 should
be carried out; the survey should be deeper than the $ROSAT$ ones so
that, if present, an X-ray faint population of cluster members could be
detected; {\bf II.} additional photometric and spectroscopic studies of
NGC~3532 should be carried out in order to confirm cluster membership for
the optical candidates known at present and to detect still unidentified
solar-type and lower mass stars in the cluster. These studies would also
provide information on rotation for cluster members; {\bf III.} If possible,
an effort should also be done, once more low-mass cluster members are known,
to provide a definitive estimate of the cluster age using also low 
main-sequence fitting.

Besides the 15 cluster members, the X-ray survey resulted in the detection
of 13 foreground/background stars -- which is not surprising given the low
cluster galactic latitude -- and of 21 objects without any known optical
counterparts. Priority should be given to optical observations aimed at
determining the nature of these sources, and, in particular, at ascertaining
whether they are cluster members or not.

\begin{acknowledgements}
We thank the referee, Dr. F. Verbunt, for useful comments and suggestions.
\end{acknowledgements}

\end{document}